\documentclass[twocolumn,preprintnumbers,amsmath,amssymb]{revtex4}
\usepackage{amsmath}    
\usepackage[dvipdfmx]{graphicx}     
\usepackage{verbatim}   
\usepackage{color}      
\usepackage{bm}
\usepackage{subfigure}  
\usepackage{hyperref}   
\usepackage[normalem]{ulem} 
\usepackage{siunitx} 

\begin{document}
\title{Geometry-driven collective ordering of bacterial vortices}
\author{Kazusa Beppu$^{1}$, Ziane Izri$^1$, Jun Gohya$^{1}$, Kanta Eto$^2$, Masatoshi Ichikawa$^2$, Yusuke T. Maeda$^{1,*}$}
\affiliation{$^1$Department of Physics, Kyushu University, Motooka 744, Fukuoka 819-0395, Japan\\ $^2$Department of Physics and Astronomy, Kyoto University, Kitashirakawa, Kyoto 606-8502, Japan}

\date{\today}

\begin{abstract}
Controlling the phases of matter is a challenge that spans from condensed materials to biological systems. Here, by imposing a geometric boundary condition, we study controlled collective motion of \textit{Escherichia coli} bacteria. A circular microwell isolates a rectified vortex from disordered vortices masked in bulk. For a doublet of microwells, two vortices emerge but their spinning directions show transition from parallel to anti-parallel. A Vicsek-like model for confined self-propelled particles gives the point where two spinning patterns occur in equal probability and one geometric quantity governs the transition as seen in experiments. This mechanism shapes rich patterns including chiral configurations in a quadruplet of microwells, thus revealing a design principle of active vortices.
\end{abstract}
\maketitle

\textit{Introduction:} In nature, collective rotational motion organized by motile elements is ubiquitous across scales, from motor proteins\cite{leibler}\cite{bausch}\cite{sumino}, flagellated sperms\cite{howard}, to the development of social amoeba cells\cite{mcnally}. Understanding the mechanism by which their motions are organized into the ordered patterns from flocking and propagating bands to a lattice of vortices is a central subject in the emerging field of active matter physics\cite{vicsek}\cite{chayes}\cite{ramaswamy}\cite{marchetti}\cite{yeomans1}\cite{petroff}. In particular, the method of controlling patterns has attracted considerable interest due to its potential in exploiting the underlying mechanism as a universal feature. As for bacteria in a quasi two-dimensional plane, dense bacterial suspensions show mesoscale collective motions in which the jets and vortices results in turbulent-like state\cite{yeomans2}. Moreover, hidden but weakly synchronized rotation appears at higher density of bacteria\cite{chate2}, implying that rotational motion may present in common. When swimming bacteria are confined in a circular space, a rotational mode of vortex arises from the guiding interaction between the bacteria and the wall\cite{goldstein}\cite{wioland1}\cite{lushi}. With the accumulated knowledge about the confinement-induced vortex of active matters, from vibrated rigid bodies\cite{nossal}\cite{chate} to colloidal rollers\cite{bartolo}, it is now apparent that a promising mean of controlling their ordered phases lies in the setting of the boundaries\cite{simmchen}\cite{gompper}.

The lattice of vortex is a conceptual basis for the description of phases of matter from magnets to superfluids and superconductors\cite{ketterle}\cite{boni}\cite{karpinsky}. This concept can impact active matters, because the correlation or frustration with defined interactions can be constructed even for bacteria. It has been reported that the ferromagnetic lattice of bacterial vortices (uniform rotational directions) or anti-ferromagnetic one (alternate clockwise and counter-clockwise rotations) was constructed in a chamber, in which vortices interact with neighbors via advection of bacteria through channels\cite{wioland2}. However, the method of coupling the vortices is actually not limited to indirect advection: vortices can be directly collided by imposing designed boundary based on geometric quantities. Hence, the geometry-based approach, by which one can control the exclusive interaction between bacterial vortices, is needed to elucidate the ordered phases and their transitions.

In this paper, we investigate the ordered phases of bacterial vortices inside microwells with designed geometries. We found that a single vortex and a doublet of vortices could be formed as the organized patterns under confinement. The pairing of vortices is classified into two distinct phases: the first one is ferromagnetic vortices (FMV) in which both vortices rotate in the same direction; the second one, anti-ferromagnetic vortices (AFMV), has the vortices rotate in opposite directions. The transition from FMV to AFMV occurs when the geometry of boundary satisfies a certain condition. Theoretical model for self-propelled particles with polar interaction in merging vortices is considered to explain the observed transition, and the predicted transition point is exactly consistent with the experiments. This new approach highlights a design principle of active vortices that could be relevant to a broad class of active matters.

\textit{Materials and Methods:}    Bacteria \textit{Escherichia coli} RP4979 strain, which was deficient of tumbling ability, was used in this study\cite{supplement}. Volume fraction of bacterial suspension was increased to 20\% to induce collective motion. To attain controlled shape of boundary, SU-8 pattern on a flat surface of Si wafer was fabricated and then used as a mold for polydimethyl siloxane (PDMS) microwells. \SI{0.5}{\micro\liter} of dense bacterial suspension was spotted on a surface treated glass slide and enclosed by PDMS microwells from the top. Typical thickness of microwells is \SI{20}{\micro\meter}. We then recorded bacterial motion in an inverted microscope at 30 frames per sec with a CCD camera for 10 sec. Velocity field of bacterial motion was acquired by PIV analysis.

\begin{figure}[t]
 \begin{center}
  \includegraphics[width=80mm]{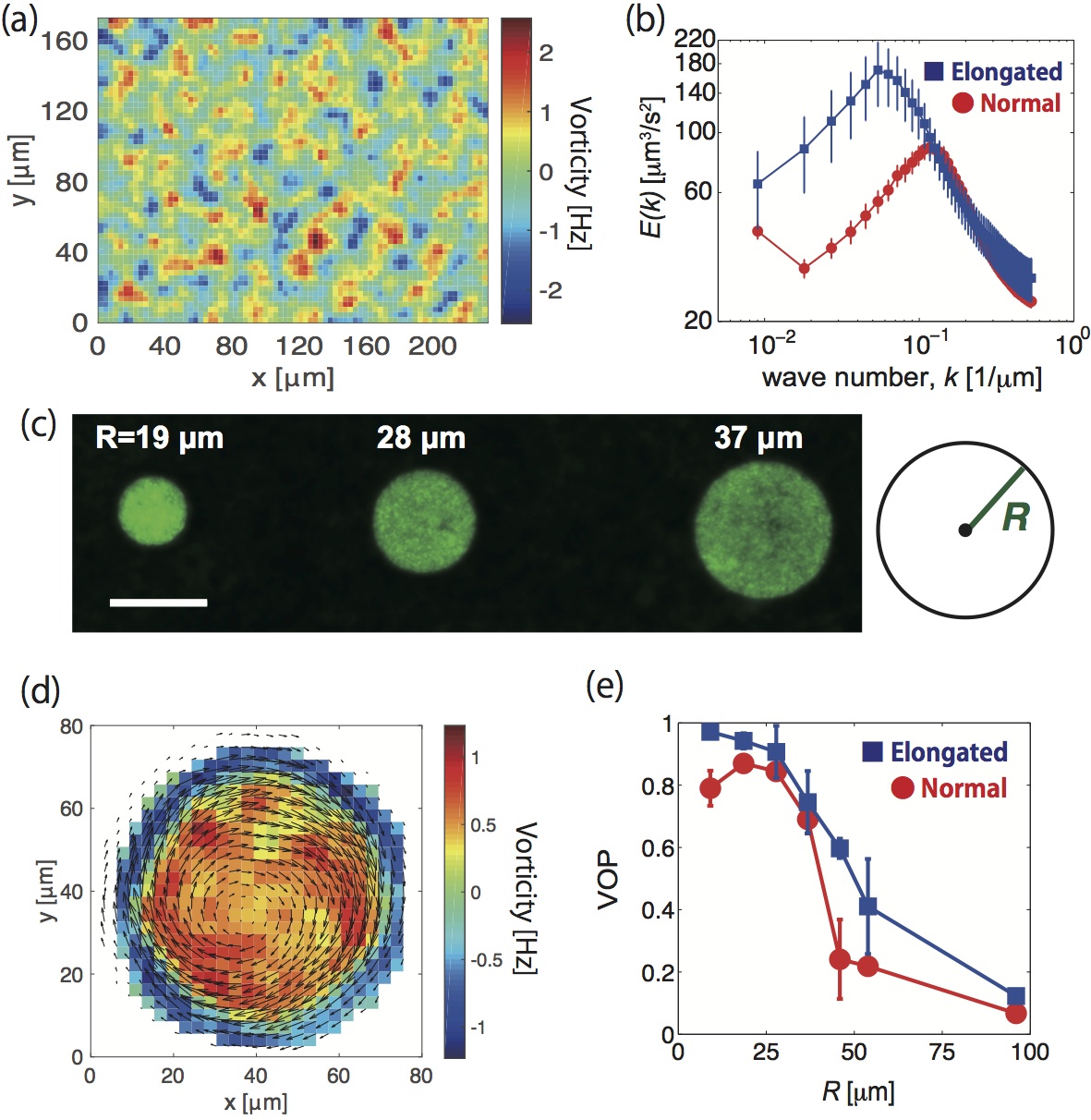}
 \end{center}
 \caption{(a) Vorticity map of the disordered pattern of RP4979 bacteria in free boundary. (b) Energy spectrum of disordered mesoscale-turbulence. Characteristic length scale $l^{*}$ is \SI{25.1}{\micro\meter} for bacteria without cephalexin (CEP) treatment (red) while it is \SI{45.2}{\micro\meter} for bacteria treated with CEP (blue). (c) Circular microwells filled with bacteria (left) and the schematic design (right). Scale bar: \SI{50}{\micro\meter}. (d) Vorticity map superposed with velocity field for CEP treated bacteria in a circular microwell of $R$=\SI{37}{\micro\meter}. (e) Vortex order parameter (VOP) for circular shape as a function of $R$. We plotted two curves of VOP calculated from normal bacteria without CEP (red) and elongated bacteria treated with CEP (blue). }\label{fig1}
\end{figure}

\textit{Results:}  When bacteria RP4979 swarm on two-dimensional plane, their collective motion is disordered (FIG. 1(a)). However, the energy spectrum exhibits a bell-shaped distribution with a long-tail, thus showing a peak at wave number $k^*$, so that the disordered vortices with a characteristic diameter, $l^{*}=\pi/k^{*}$ are present, although they are apparently hidden in turbulent-like motion (FIG. 1(b))\cite{supplement}. To test whether a single vortex can be isolated by imposing a boundary condition, we construct microwells of circular boundary for various radius $R$, as a simplest model (FIG. 1(c)). FIG. 1(d) shows a typical example of a velocity field $\bm{v}(r,\theta)$ of bacterial motion in a circular microwell, in which a single vortex is formed. Vorticity $\bm{\nabla}\times\bm{v}$ exhibits positive (or negative) values around the center of the microwell, while close to the wall at $r=R$, it shows opposite values because of the decay of the velocity nearby the boundary. 

What one needs to know in order to control collective motion is the upper limit for selecting single vortex. We in turn examined the size-dependence of vortex formation in circular microwells with various $R$ from 10 to \SI{100}{\micro\meter}. Vortex order parameter (VOP) of single vortex is defined as $\frac{1}{1-2/\pi} \bigl(\frac{\sum_i \vert \bm{v}_i\cdot \bm{t}_i \vert}{\sum_i \| \bm{v}_i \|} - \frac{2}{\pi} \bigr)$ \cite{goldstein}\cite{wioland1} where $i$ represents the index of sites in the microwell and $\bm{t}_i$ is the unit orthoradial vector at site $i$. VOP was employed to classify either single vortex (VOP=1) or disordered motion (VOP=0). FIG. 1(e) (red) exhibits that VOP is between 0.7 to 0.9 for $R$=10 to \SI{37}{\micro\meter} whereas VOP drops to 0.2 for $R\geq$\SI{46}{\micro\meter}. It means that $R$=\SI{37}{\micro\meter} is close to the limit where rotational velocity correlation persists over the confined space. This size is comparable to $l^{*}\approx$\SI{25}{\micro\meter}, implying that additional vortices arise in the same microwell as $R$ becomes much larger than $l^*$. 

The correlation between $R$ and $l^*$ was further examined by using bacteria having a longer body. It is assumed that a longer rod-shaped body may enhance the local alignment of bacteria and in turn alters the size of vortices in mesoscale turbulence\cite{yeomans1}. For \textit{E.coli} bacteria, a short exposure to a cell-division inhibitor (cephalexin, CEP) makes them become more elongated: the average length is \SI{14.5}{\micro\meter} for elongated bacteria but the one for untreated bacteria is  \SI{8.3}{\micro\meter}\cite{supplement}\cite{sano}. The elongation of the bacteria shifted the peak in the energy spectrum to $l^{*}=45.3\pm$\SI{0.1}{\micro\meter} (FIG. 1(b), blue). These bacteria also show a single isolated vortex in circular microwells as we expected, and then sustained higher VOP$\geq$0.5 for $R$=19 to \SI{46}{\micro\meter}, which was, once again, comparable to $l^{*}$ (FIG. 1(e), blue). Thus, we demonstrate that the range of vortex formation is prolonged by altering correlation length scale in velocity, indicating that the confinement by $R\lesssim l^{*}$ is required to host a single vortex.
We found that the number of vortices with counter-clockwise rotation is 28 out of 48 samples, which consist of 14 for normal RP4979 bacteria and 34 for elongated bacteria, selected by VOP$\geq$0.8. The proportion of clockwise and counter-clockwise is 58\%, meaning that the rotational direction of single vortices is symmetric. We also note that the layer of counter-rotation against a confined vortex (negative slip velocity \cite{wioland1}) is observed on occasion at 5 samples out of 48 in total (vortices with large VOP$\geq$0.8). Most of bacterial vortices observed here have either no-slip or positive small slip boundary, implying that counter-rotating action appears to be secondary effect.

\begin{figure}[t]
 \begin{center}
  \includegraphics[width=90mm]{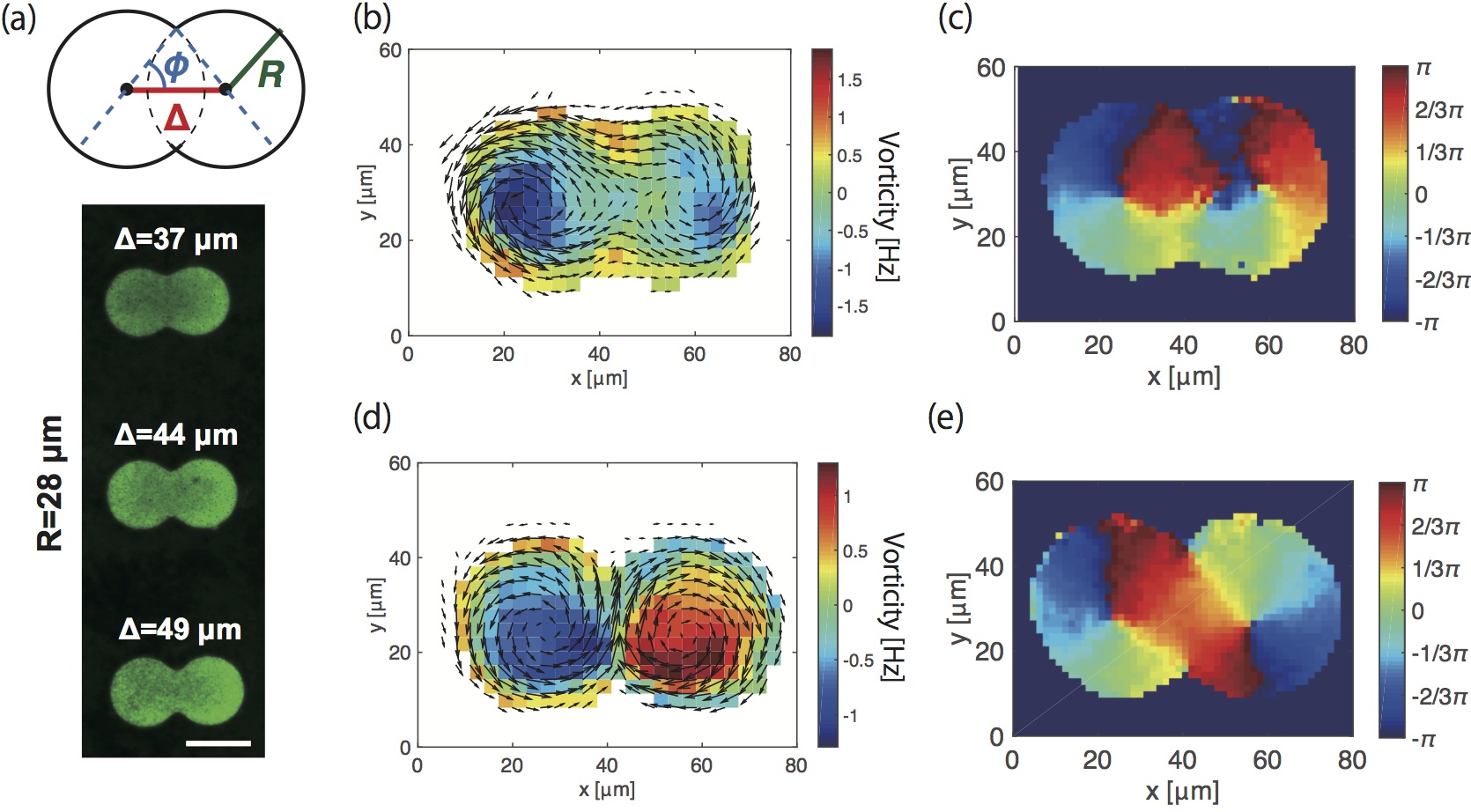}
 \end{center}
 \caption{Pattern formation of vortex pairing in a doublet of circular microwells (Dcm). (a) Representative pictures and the schematic description of a Dcm of $R$=\SI{28}{\micro\meter}. Its geometry can be given by two geometric quantities $\Delta$ and $R$. These quantities are also linked by the elevation angle $\phi$ as $\cos \phi =\Delta/(2R)$. Scale bar is \SI{50}{\micro\meter}. (b) Velocity field merged with vorticity map of a Dcm of $R$=\SI{19}{\micro\meter} and $\Delta$=\SI{25}{\micro\meter}. (c) Orientation map of velocity field corresponding to (b). (d) Velocity field and vorticity map of a Dcm of $R$=\SI{19}{\micro\meter} and $\Delta$=\SI{31}{\micro\meter}. (e) Orientation map of velocity field corresponding to (d).}\label{fig2}
\end{figure}
 
Let us now consider two confined vortices interacting with one another, and how the geometry drives their behavior. We construct a doublet of circular microwells (Dcm) with two identical and overlapping circles (FIG. 2(a)). Such a geometry is chosen because it can be simply drawn with two geometric quantities: the radius of circles $R$ and the distance between their centers $\Delta$. In fact, those quantities are also linked by the angle of elevation $\phi$ with regard to the horizontal axis, as $\cos \phi=\Delta/(2R)$. Hence, a doublet shape offers explicit definition of boundary to analyze geometry-induced phenomena. FIG. 2 shows that bacterial motion was organized into two vortices whose rotational directions are either identical or opposite. When the two circles have a large overlapping area, i.e. small $\Delta$, a single vortex is no longer sustained but instead two vortices with the same direction of rotation appear (FIGs. 2(b) and (c)). We name this pattern as ferromagnetic vortices (FMV) after the spinnings in parallel. However, as $\Delta$ is increased, the spinning of vortices becomes opposite, with a pair of clockwise and counter-clockwise rotations (FIGs. 2(d) and (e)). This configuration of vortices is named as anti-ferromagnetic vortices (AFMV) after the spinnings in anti-parallel. Both FMV and AFMV patterns are stable for \SI{10}{s} during experimental observations and spontaneous transition from one to the other is not observed. Changing $\Delta$ is therefore a key parameter to determine one favorable vortex pairing out of two possible patterns. 

The transition from FMV to AFMV occurs at $\Delta=\Delta_c$ but $\Delta_c$ presents a dependence on the circle radius $R$: the onset of AFMV pattern is at $\Delta_c$=\SI{26}{\micro\meter} for $R$=\SI{19}{\micro\meter} but it is shifted to $\Delta_c$=\SI{53}{\micro\meter} for larger $R$=\SI{37}{\micro\meter}. What geometric rule associated with $\Delta$ and $R$ characterizes the transition? Interestingly, when we display the occurrence of FMV and AFMV in the ratio of $\Delta/R$, which is linked to $\cos \phi$, the transition point between these two patterns collapses to a single horizontal line, i.e. $\Delta_c/R\approx1.4$, even for $R$ varying from 19 to \SI{37}{\micro\meter} (FIG. 3(a)).

\begin{figure}[t]
 \begin{center}
   \includegraphics[width=80mm]{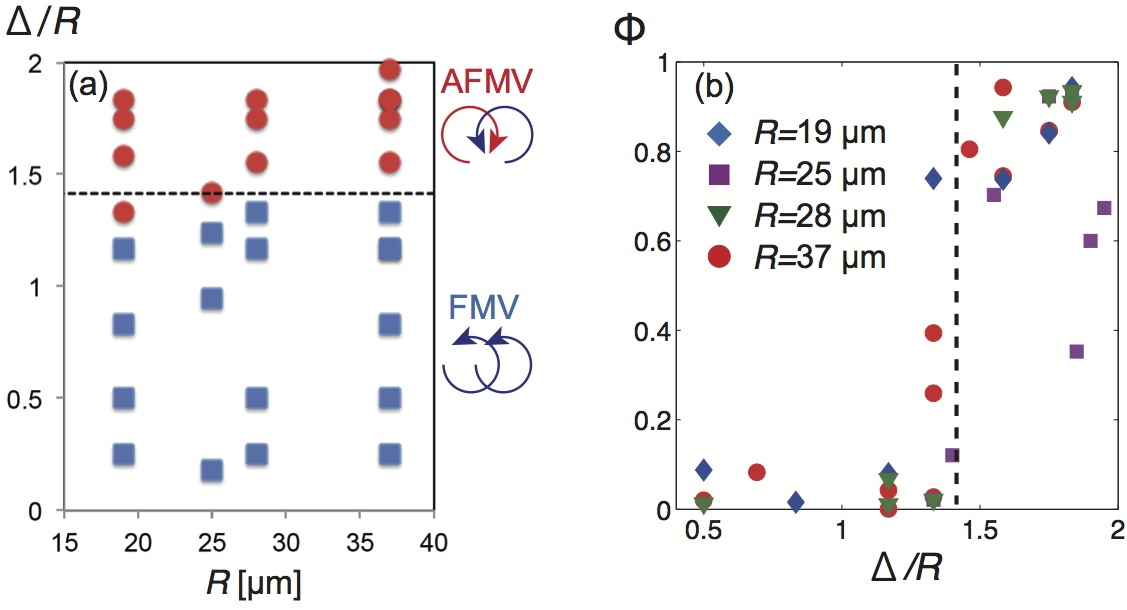}
   \end{center}
 \caption{Transition of vortices configuration. (a) Phase diagram of vortices pattern plotted on $\Delta/R$ - $R$ plane. Vortices switch from FMV (blue) to AFMV (red) across $\Delta\approx1.4R$. Dashed line is transition point theoretically predicted by Eq. (5). (b) Order parameter $\Phi$ of AFMV pattern is plotted in $\Delta/R$ for varying $R$=\SI{19}{\micro\meter}(diamond), \SI{25}{\micro\meter}(square), \SI{28}{\micro\meter}(triangle), \SI{37}{\micro\meter}(circle). $\Phi$ rises sharply at $\Delta_c/R\approx$1.4.}\label{fig3}
\end{figure}

 To resolve the relation shown above, the order parameter of AFMV pattern, $\Phi$, is defined as 
\begin{equation}
\Phi=\vert \langle \bm{p}_i \cdot \bm{u}_i \rangle \vert
\end{equation}
where $\bm{p}_i$ is the orientational map measured experimentally at site $i$ in a Dcm and $\bm{u}_i$ is the expected orientation of velocity of AFMV calculated numerically at corresponding site\cite{supplement}. We take ensemble average $\langle \cdot \rangle$ over all possible sites $i$ inside a Dcm. $\Phi$ reaches 1 for AFMV while it goes down to 0 for FMV due to opposite sign of the product with one counter-rotating vortex. As a common feature among all sizes of microwell, FIG. 3(b) shows that $\Phi$ sharply increases from 0 to nearly 0.9 at around $\Delta_c/R$=1.4, yet it is independent of $R$. This finding implies that the ratio $\Delta/R$ is the proper parameter to control vortex pairing. 

To elucidate the mechanism of the transition from FMV to AFMV, we decide to analyze a Vicsek-like model based on confined self-propelled particles\cite{vicsek}. Point-like particles at position $\bm{x}_m$ move at speed $v_0$ along their heading $\theta_m(t)$, i.e. $\dot{\bm{x}}_m$=$v_0 \bm{e}(\theta_{m})$ where $\bm{e}(\theta_{m})$=$(\cos\theta_m, \sin\theta_m)$. Headings of particles evolve following the relation $\dot{\theta}_m = - \gamma \frac{\partial U}{\partial \theta_m} + \eta_m(t)$ where $U$ is the potential describing the polar alignment interaction with neighbors and it is defined as $U(\bm{x}_m,\theta_m) = - \sum_{\vert \bm{x}_m-\bm{x}_n \vert < \epsilon} \cos(\theta_m - \theta_n)$. $\epsilon$ determines the effective radius of particle interaction. $\eta(t)$ is Gaussian white noise $\langle \eta_m(t) \eta_n(t') \rangle$=$2D\delta_{mn}\delta(t-t')$ where $\delta_{mn}$ and $\delta(t-t')$ are Dirac delta functions. Moreover, we take the effect of the boundary as a nematic interaction with motile particles and set a Dcm boundary condition with a geometric quantity $\phi$. 
Assuming a homogeneous spatial distribution of the particles, necessary for mean-field approximation\cite{peruani}, the Fokker-Planck equation expressing the probability distribution $P(\theta)$ of a particle having a heading $\theta$ in the middle of the Dcm is
\begin{equation}
\frac{\partial P}{\partial t} = D \frac{\partial^2 P}{\partial \theta^2} + \gamma \frac{\partial}{\partial \theta} \Bigl( \int^{\pi}_{-\pi} \sin(\theta - \theta') \bar{P}(\theta',t;\phi)d\theta' P(\theta,t;\phi) \Bigr)
\end{equation}
where $\bar{P}(\theta',t;\phi)$ is the probability of heading $\theta'$ from either left or right microwell in a doublet described by $\phi$\cite{supplement}. Because particles start interacting at the tip in the middle of a Dcm, $\bar{P}(\theta';\phi)$ from left microwell is given either by $\bar{P}(\theta'; \phi)=\delta(\theta' - \pi/2 + \phi)$ (counter-clockwise rotation) or $\bar{P}(\theta'; \phi)=\delta(\theta' + \pi/2 + \phi)$ (clockwise rotation) with low noise limit where bacteria move along the boundary (FIG. 4(a))\cite{supplement}. Symmetrically, the probability of having an orientation $\theta'$ for particles at the tip, coming from the right circle is $\bar{P}(\theta'; \phi)=\delta(\theta' - \pi/2 - \phi)$ (clockwise rotation) or $\bar{P}(\theta'; \phi)=\delta(\theta' + \pi/2 - \phi)$ (counter-clockwise rotation). As vortex pair forms respectively FMV and AFMV patterns, one derives the solution of Eq. (2) at the steady state
\begin{eqnarray}
P^{FMV}(\theta;\phi) = \frac{\exp(\frac{2\gamma}{D} \cos \theta \sin \phi)}{2\pi I_0(\frac{\gamma}{D}\sin \phi)}\\
P^{AFMV}(\theta;\phi) = \frac{\exp(\frac{2\gamma}{D} \sin \theta \cos \phi)}{2\pi I_0(\frac{\gamma}{D} \cos \phi)}
\end{eqnarray}
Hence, at the condition where $\sin \phi_c = \cos \phi_c$, i.e. $\phi_c = \pi/4$, $P^{FMV}(\theta=0;\phi=\pi/4)=P^{AFMV}(\theta=\pi/2;\phi=\pi/4)$. In other words, the geometry of $\phi<\pi/4$ (or $\phi > \pi/4$) selects AFMV pattern (or FMV pattern) (FIG. 4(b)). This symmetry allows us to obtain $\Delta_c$ as
\begin{equation}
\frac{\Delta_c}{2R} = \cos\Bigl(\frac{\pi}{4}\Bigr).
\end{equation}
Eq. (5) immediately leads to $\Delta_c/R=\sqrt{2}\approx1.41$, which exactly agrees with the experimental results. This fact indicates that the tip in the central area plays a crucial role for controlling the preferred direction of the alignment and in turn determines one vortex pairing as either FMV or AFMV.
\begin{figure}[t]
 \begin{center}
  \includegraphics[width=65mm]{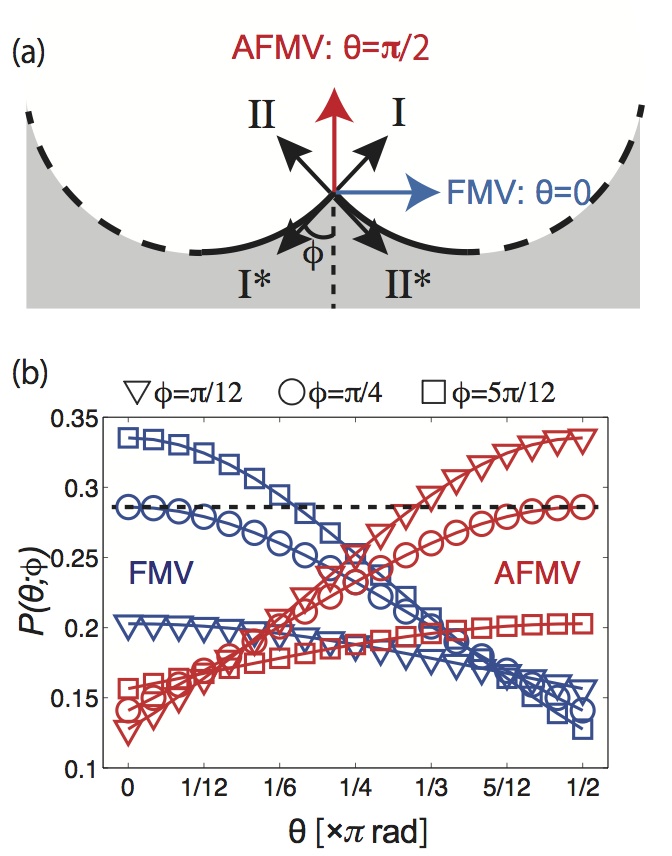}
   \end{center}
 \caption{Theoretical model for the transition of vortex pairing. (a) Schematic of particle motion at the vicinity of the tip between two circles. At the tip, the particles in left circle go either in the direction of $I$ or $I^*$, while the ones in the right circle go either in the direction of $I\hspace{-.1em}I$ or $I\hspace{-.1em}I^*$. Alignment at the tip decides whether two vortices make a FMV or AFMV pattern. (b) Probability distributions of $P^{FMV}$ (blue) and $P^{AFMV}$ (red) are plotted as functions of $\theta$ at $\phi$=$\pi/12$, $\pi/4$, and $5\pi/12$. Here we take $2\gamma/D=1$. Dashed black line indicates the equal probability of $P^{FMV}$ and $P^{AFMV}$ at $\phi=\pi/4$.}\label{fig4}
\end{figure}

It is worth mentioning about hydrodynamic effect in our observation. Bacteria are regarded as a force-dipole\cite{ramaswamy} and driven-fluid flow leads to disordered motion such as mesoscale turbulence. In our study, this effect may have little amplitude because the radius of circular microwell is smaller than $l^*$ so as to avoid loss of alignment correlation over distance. In addition, it has been reported that swimming bacteria are trapped at curved walls\cite{gompper}\cite{sipos} whereas bacteria uniformly distribute with less heterogeneity in our microwells (FIG. 2(a) and FIG. 3(a)), implying that hydrodynamic trapping may be suppressed due to high density of bacteria.

\begin{figure}[t]
 \begin{center}
  \includegraphics[width=80mm]{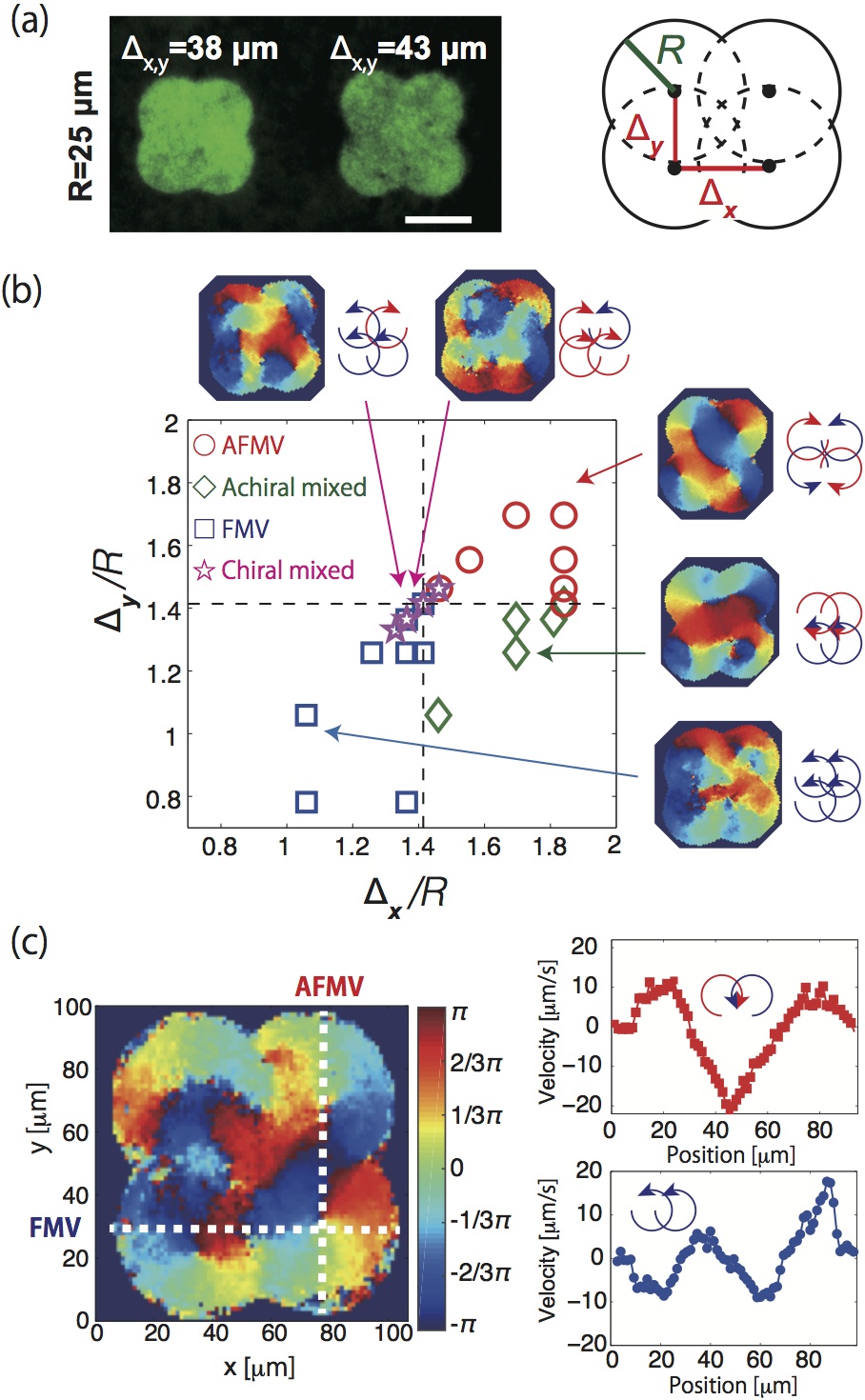}
 \end{center}
 \caption{Design of complex patterns of vortices. (a) Quadruplet of microwells (Qcm) and its schematic design. Scale bar: \SI{50}{\micro\meter}. (b) Phase diagram of collective vortices. The mixed states of FMV and AFMV with achiral symmetry occur in asymmetric quadruplets (green diamond). The mixed configurations with chiral symmetry are found only close to transition point $\Delta_{x,c}/R=\Delta_{y,c}/R=\sqrt{2}$ (purple star). (c) (left) Orientation map of vortices in a Qcm of $R$=\SI{25}{\micro\meter} and $\Delta_x/R=\Delta_y/R=1.33$. (right) Line scan of normal velocity over dotted diagonal lines on orientation map.}\label{fig5}
\end{figure}

Finally, in order to test the predictability of our theoretical model for more complex boundary, we examined ordered pattern of vortices in a quadruplet of circular microwells (Qcm)(FIG. 5(a)). On the one hand, we found that all four vortices rotate in the same direction (FMV) when both $\Delta_x/R \leq \sqrt{2}$ and $\Delta_y/R \leq \sqrt{2}$ were satisfied. On the other hand, the vortex pairing becomes AFMV when both $\Delta_x/R \geq \sqrt{2}$ and $\Delta_y/R \geq \sqrt{2}$ (FIG. 5(b)). By imposing not only a symmetric boundary $\Delta_x/R=\Delta_y/R$ but also an asymmetric one such as $\Delta_x/R\geq\sqrt{2}$ and $\Delta_y/R\leq\sqrt{2}$, we find a mixed configuration of FMV and AFMV with achiral symmetry (FIG. 5(b), green).  

Strikingly, despite the absence of asymmetry in geometry, FMV and AFMV coexist in the region close to the transition point, e.g. $\Delta_x/R=\Delta_y/R=1.33$ (FIG. 5(c)). Eqs. (3)-(4) account for this coexistence because the probabilities of FMV and AFMV become comparable as $\phi$ is close to $\pi/4$. The exclusive interaction among vortices no longer discriminates between FMV and AFMV, and the 
ordered patterns can also exhibit chiral configurations (FIG. 5(b), purple). Thus, we conclude that theoretical model constructed here can be applied even to miscellaneous geometries so that it draws a design principle for bacterial vortices pairing.

\textit{Discussion:}  We have studied collective ordering of bacterial vortices in a confined space with defined geometries. For bacterial vortices, the probability of orientation of bacterial motion from the tip can be controlled by geometry, e.g. FMV pattern is permitted when $\Delta/R\leq\sqrt{2}$. Our data show that one geometric parameter $\Delta/R$ is sufficient to control the vortex pairing. This finding allows one to consider geometry as a powerful mean to dictate spatial orderings of active vortices. For original Vicsek model, the point-like particles with less fluctuation are depleted from the center of confined space and then results in large density heterogeneity while, for our experiment, the density of bacteria is almost uniform inside microwells. Eqs (3) to (5) derived from mean-field approximation explain the transition point but further investigation, e.g. self-propelled particles with excluded volume, remained as a future work to understand pattern formation of FMV and AFMV. We also found typical size of vortices increases in elongated bacteria and this empirical observation was employed in order to examine the transition for various sizes of confinement. The underlying mechanism of this size-dependence is left as a subject for future study. Thanks to its simplicity, our theoretical model may provide versatile protocol for not only bacteria but also active cytoskeletons\cite{sanchez}. In that case, an elucidated mechanism can be in turn used to rationally design an autonomous transporter of small objects over a long distance\cite{dogic}. To realize such an isothermal engine, the extended analysis of frustrated vortices in a triplet of circular microwells appears to be promising. Finally, the transition of vortex pairings may provide new insight into the universality for the ordered phases of matter; it has analogous symmetry to the transition between type {I} and {II} in a superconductor\cite{karpinsky}. In the future, a comprehensive exploration of collectively ordered vortices will be developed for uncovering design principles that hold in wide classes of matter from colloidal rollers\cite{bartolo} and cytoskeletons\cite{miyazaki} to quantum systems\cite{karpinsky}. 

\textit{Acknowledgements}  We thank I. Kawagishi for providing RP4979 strain. This work was supported by PRESTO (No.11103355, JPMJPR11A4) from JST, KAKENHI (No. 16H00805: "Synergy of Fluctuation and Structure") from MEXT, and HFSP (RGP0037/2015).

\textit{corresponding address:}
kazu.beppu@phys.kyushu-u.ac.jp or ymaeda@phys.kyushu-u.ac.jp

\newpage

\section*{Supplementary material\\Geometry-driven collective ordering\\ of bacterial vortices\\K. Beppu, Z. Izri, J. Gohya, K. Eto, M. Ichikawa and Y.T. Maeda}

$\textbf{Bacterial culture}$

  Bacteria \textit{Escherichia coli} RP4979 strain that was deficient of tumbling ability due to the lack of \textit{cheY} gene was used. RP4979 bacterial strain was transformed with a plasmid DNA that encodes constitutive expression of YFP protein. YFP expression allows to test spatial homogeneity of bacteria in microwells. We inoculate single colony of RP4979 in LB medium (NaCl 10 g/L, Tryptone 10 g/L, Yeast extract 5 g/L, autoclaved at \SI{120}{\celsius} for 20 min) with selective antibiotics (\SI{25}{\micro\gram}/mL chrolamphenicol) and then have bacteria grow at \SI{37}{\celsius} with shaking 150 rpm. The overnight culture was diluted by a factor of 100 in T-broth (NaCl 10 g/L, Tryptone 10 g/L, autoclaved at \SI{120}{\celsius} for 20 min) of 50 mL with selective antibiotics and diluted culture was incubated at \SI{30}{\celsius} with shaking at 150 rpm until bacterial density reached O.D.$_{600}$=0.4. Grown culture was centrifuged at 3000 rpm for 10 min and the supernatant was removed carefully in order to increase the volume fraction of bacteria to 20\%(v/v). 

In addition, we used elongated bacteria that were prepared by the exposure to \SI{20}{\micro\gram}/mL of cephalexin (CEP), an inhibitor of bacterial cell division, for 1 hour just before the end of cultivation. The elongated bacteria tends to form larger vortex as shown in FIG. 1. Bacteria used in this study were transformed with a plasmid encoding YFP protein and its YFP fluorescence allows to measure the size of individual bacteria accurately. We analyzed the size of individual bacteria by conventional image processing and the averaged size of long axis is linearly increased with the duration of CEP treatment. The velocity of bacteria after CEP treatments for various exposure durations was measured as the displacement of the center of mass. The speed of bacterial motion is comparable to that of untreated bacteria. However, we found that the maximum speed in the velocity field after PIV is 9.4$\pm$\SI{2.0}{\micro\meter}/s for CEP treated bacteria, which is comparable or slightly faster than the maximum speed of 8.4$\pm$\SI{0.2}{\micro\meter}/s in PIV for untreated bacteria. This difference in velocity may result from the hypothetical correlation among the size of bacteria, the size of vortices, and the alignment of bacteria but it is out of the focus of this study.

$ $

$\textbf{Microfabrication}$

　We used SU-8 3025 photoresist (Microchem) for all the photolithographies necessary in this study. Chromium masks (MITANI micronics, Japan) were used to print patterns in the photoresist during an exposure to UV light in a mask-aligner (MA-100, MIKASA, Japan). Molds of poly-dimetyhl siloxane (PDMS) microwells were cured on the surface of silicon wafers, while the surface of SU-8 patterns was smoothed by coating with CYTOP, a fluorinated coating agent (Asahi glass, Japan). PDMS elastomer was cast on top of the patterned SU-8 mold and cured at \SI{70}{\celsius} for 1 hour. The hardened PDMS was cut with a scalpel and the patterned surface was coated with MPC polymer (Lipidure, Nichiyu Coop., Japan) and heated for 1 hour at \SI{50}{\celsius},  which increased its hydrophilicity, to avoid non-specific adhesion of bacteria. The thickness of the PDMS microwells was measured by laser scanning surface profiler (LT-9000, Keyence, Japan) and it was about \SI{20}{\micro\meter}. The glass cover slips used as the bottom of the microfluidic chips were also coated with MPC according to the same recipe in order to avoid the non-specific adhesion of bacterial bodies. 

$ $

$\textbf{Image acquisition and processing}$

\SI{0.5}{\micro\liter}  of dense bacterial suspension was put onto the MPC-coated coverslip. Thereafter, MPC-coated PDMS microwells were placed on top of the droplet and then pressed to enclose bacterial suspension. Bright-field optical imaging and video-microscopy were performed using an inverted microscope (IX73, Olympus). We recorded swarming motion of bacteria at a rate of 30 fps with a CCD camera (DMK23G445, Imaging Source) controlled by custom made LabVIEW program. The experimental data was acquired within \SI{30}{min} after preparing dense bacterial suspension in order to avoid the proliferation (doubling time is about \SI{1.5}{h} in T-broth) and to use fresh bacteria without losing motility. Velocity fields of bacterial swarm were obtained by PIV with Wiener filter method using PIVlab based on MATLAB software. Acquired velocity fields were further smoothed by averaging over 30 frames that correspond to \SI{1}{s}. To analyze disordered state of bacterial vortices as shown in FIG. 1(a), we calculated energy spectrum $E(k)$ of two-dimensional space that indicates the kinetic energy at the wavenumber $k=2\pi/r$. Two-dimensional energy spectrum can be obtained by Fourier transform of two-point velocity correlation function as $E(k)=\frac{k}{2 \pi} \int d^2{r'} e^{-i {k}\cdot{r'}} \langle {v}({r},t) \cdot {v}({r}+{r'},t)\rangle$ where ${r'}$ is the distance between two arbitrary points for the calculation of velocity correlation function at the same time point $t$\cite{yeomans2}. 

$ $ 

Here we consider self-propelling point-like particles that can interact through a potential $U$ of polar alignment. The state of particle $m$ at time $t$ is represented by two variables, its coordinate $\bm{x}_m(t)$ and its orientational angle of motion $\theta_m(t)$. Particles align their direction of motion through $\partial_{\theta} U(\bm{x}_m,\theta_m)$ and their relaxation coefficient is given by $\gamma$. Hence, the evolution of $\bm{x}_m(t)$ and $\theta_m(t)$ belong to a Vicsek-like model as follow:

\begin{equation}
\dot{\theta}_m = - \gamma \frac{\partial U}{\partial \theta_m} + \eta_m(t)
\end{equation}
where $\eta_m(t)$ is random noise, which means that the direction of motion of the particles is random at infinite dilution limit, and its correlation satisfies $\langle \eta_m(t) \rangle$=0, $\langle \eta_m(t) \eta_n(t') \rangle$=$2D \delta_{mn}\delta(t-t')$ where $\delta_{mn}$ and $\delta{t}$ is Dirac delta function. $D$ is the diffusion constant in rotational direction, which is related to noise strength. 

In addition, for two-dimensional coordinate,

\begin{equation}
\dot{\bm{x}}_m = v_0 \bm{e}(\theta_{m})
\end{equation}
where $\bm{e}(\theta_{m})$ is unit vector of velocity defined as $\bm{e}(\theta_{m})$=$(\cos\theta_m, \sin\theta_m)$. We can easily find that particles move at a constant speed $v_0$ while fluctuation is involved in rotational direction alone. 

The alignment of velocity vector is based on polar interaction and hence the potential $U(\bm{x}_m,\theta_m)$ is 

\begin{equation}
U(\bm{x}_m,\theta_m) = - \sum_{\vert \bm{r}_{mn} \vert < \epsilon} \cos(\theta_m - \theta_n)
\end{equation}
where $\bm{r}_{mn}=\bm{x}_m-\bm{x}_n$ and  $\epsilon$ is the effective radius of particle interaction.

$ $ 
We consider a distribution of particles showing homogeneous spatial distribution. Namely, probability distribution is a function of $\theta$ and $t$. For this case, Fokker-Planck equation of the point-like particles is given by
\begin{equation}
\frac{\partial P}{\partial t} = D \frac{\partial^2 P}{\partial \theta^2} + \gamma \frac{\partial}{\partial \theta} \Bigl( \int^{\pi}_{-\pi} \sin(\theta - \theta')P(\theta',t)d\theta' P(\theta,t) \Bigr)
\end{equation}
where $P(\theta,t;\phi)$ is the probability distribution of particles heading $\theta$ at time $t$. The focus of this theoretical analysis is to find analytical solution that can account for the transition from FMV to AFMV observed in experiment. In that sense, what we need to consider is the interaction of particles from left or right circles in a doublet microwell defined by geometric constant $\phi$. As for this case, the Fokker-Planck equation can be expressed by
\begin{equation}
\frac{\partial P}{\partial t} = D \frac{\partial^2 P}{\partial \theta^2} + \gamma \frac{\partial}{\partial \theta} \Bigl( \int^{\pi}_{-\pi} \sin(\theta - \theta')\bar{P}(\theta',t;\phi)d\theta' P(\theta,t;\phi) \Bigr)
\end{equation}
where $\bar{P}(\theta',t;\phi)$ is the probability distribution of the orientation of particles $\theta'$ at the tip from either left or right circle. Hence, the probability distribution $P(\theta,t;\phi)$, meaning the orientation angle $\theta$ of particles rectified by the polar alignment at the tip, is able to be derived as the analytical solution of Eq. (4) once we get the explicit form of $\bar{P}(\theta',t;\phi)$. Therefore, we next consider the motion of particles due to the association with boundary wall to find the form of $\bar{P}(\theta',t;\phi)$. 

The interaction between motile particles $m$ and the wall $\bar{n}$ is assumed nematic. Fokker-Planck equation of the heading $\theta$ of particles associated with the boundary is given by
\begin{equation}
\frac{\partial \bar{P}}{\partial t} = \bar{D} \frac{\partial^2 \bar{P}}{\partial \theta^2} + \bar{\gamma} \frac{\partial}{\partial \theta} \Bigl( \int^{\pi}_{-\pi} \sin \bigl(2(\theta - \theta') \bigr)\bar{P}(\theta',t)d\theta' \bar{P}(\theta,t) \Bigr)
\end{equation}

\begin{equation}
\bar{U}(\bm{x}_m,\theta_m) = - \sum_{\vert \bm{r}_{m\bar{n}} \vert < \bar{\epsilon}} \cos \bigl(2(\theta_m - \theta_{\bar{n}}) \bigr)
\end{equation}
where $\bm{r}_{m\bar{n}}=\bm{x}_m-\bm{x}_{\bar{n}}$ and  $\bar{\epsilon}$ represents the range of the effective nematic interaction between a particle $m$ and a wall $\bar{n}$. We note that a vortex in a circular microwell and vortex pairing patterns (FMV and AFMV) in a doublet microwell are persistent in time within the range of measurement. This fact allows us to consider the steady state, $\partial_t P=0$ and $\partial_t \bar{P}=0$, to analyze Eqs. (5) and (6), respectively. The solution of Eq.(6) at the steady state is 
\begin{equation}
\bar{P}(\theta)=\frac{1}{2 \pi I_0(\alpha\bar{\gamma}/\bar{D})} \exp\Biggl[\frac{\alpha\bar{\gamma}}{\bar{D}}\cos2(\theta- \theta_0) \Biggr]
\end{equation}
where $\alpha=\int^{\pi}_{-\pi} \cos(2\theta)\bar{P}(\theta)d\theta$ and $I_0(x)$ is modified Bessel function of the first kind and $\theta_0$ is the tangential angle at the boundary. Close to the boundary, the nematic interaction with the wall is assumed strong enough to neglect the angular noise, so that $\bar{\gamma}/\bar{D} \rightarrow \infty$. The condition of low noise reflects the state which $\bar{P}(\theta)$ is no longer constant and thereby one can find $\alpha \neq 0$. The probability distribution is rewritten as
\begin{equation}
\lim_{\bar{\gamma}/\bar{D} \to \infty} \frac{ \exp\Bigl(\frac{\alpha \bar{\gamma}}{\bar{D}} \cos2(\theta- \theta_0) \Bigr) }{2 \pi I_0(\alpha \bar{\gamma}/\bar{D})} = \delta(\theta - \theta_0 -l\pi)
\end{equation}
where $\delta(\theta)$ is the Dirac delta function and $l$ is $0, \pm1, \pm2, \cdots$ but $\delta(\theta - \theta_0)$ and $\delta(\theta - \theta_0 - \pi)$ are taken to describe either clockwise or counter-clockwise motion along the boundary for later analysis. Thus, the explicit form of $\bar{P}(\theta'; \phi)$ can be obtained by considering the tangential direction of the curved boundary at the tip. 

As for a doublet of circular microwells (Dcm) with geometrical parameter $\phi$, given that particles enter into left microwell by either incoming or outgoing direction at the tip, the probability of particle heading $\theta'$ from left is given by

($I$) Outgoing from left microwell
\begin{equation}
\bar{P}(\theta'; \phi)=\delta(\theta' - \pi/2 + \phi),
\end{equation}
or 

($I^*$) Incoming into left microwell
\begin{equation}
\bar{P}(\theta'; \phi)=\delta(\theta' + \pi/2 + \phi).
\end{equation}
where we use the relation $\theta_0=\phi \pm \pi/2$ nearby the tip of Dcm.
The particles that move along the boundary of a doublet microwell interact close to the tip. Hence, in addition to Eqs. (10) and (11), one needs to take the bacterial motion from the right side into account so as to describe the collective motion after the association between particles coming from both left and right sides. Hence, the probability of particle heading $\theta'$ from right is given by

($I\hspace{-.1em}I$) Outgoing from right microwell
\begin{equation}
\bar{P}(\theta'; \phi)=\delta(\theta' - \pi/2 - \phi),
\end{equation}
or 

($I\hspace{-.1em}I^{*}$) Incoming into right microwell
\begin{equation}
\bar{P}(\theta'; \phi)=\delta(\theta' + \pi/2 - \phi).
\end{equation}

On the one hand, as shown in FIG. 4(a), one can assume that particles can form AFMV pattern when polar interaction of (($I$) and ($I\hspace{-.1em}I$)) or (($I^*$) and ($I\hspace{-.1em}I^{*}$)) dominantly occurs at the middle. On the other hand, FMV pattern results from polar interaction of (($I$) and ($I\hspace{-.1em}I^{*}$)) or (($I^{*}$) and ($I\hspace{-.1em}I$)) because the group of particles keep moving along boundary wall. Therefore, by taking one pair of two explicit forms $P(\theta';\phi)$ given above, one can solve Fokker-Planck equation of Eq. (10) and finally obtain the probability distribution of particle heading $\theta$ at the tip, as given in Eqs. (3) and (4) in main text.

$ $

$\textbf{Velocity of a vortex in a circular microwell}$

In this section, we derive the function of angular velocity of single vortex, ${v}_{\theta}(r)$, formed inside a circle of the radius $R$. We assume that velocity of bacterial swarming decays at the vicinity of boundary wall so that the boundary condition at $r=R$ is ${v}_{\theta}(r)$=0. However, ${v}_{\theta}(r)$ is proportional to $r$ and does not satisfy the above condition if one supposes uniform vorticity inside the circle $r<R$. To reconcile both vortex formation and the boundary condition at $r=R$, the superposition of two different vortices has to be taken into account. Indeed, FIG. 1(d) exhibits the presence of two regions with opposite vortices. Hence, the spatial distribution of vorticity inside the circle is given by

\begin{equation}
  \omega(r) = \begin{cases}
    \omega & (0 \leq r \leq R-s) \\
    - \omega \Bigl[1- \frac{(R-s)^2}{R^2} \Bigr]  & (R-s \leq r \leq R)
  \end{cases}
\end{equation}
where $R-s$ is the position we find the peak of angular velocity. By solving Laplace equation, the analytic expression of the orthoradial velocity in a circular microwell $\bm{v}(r,\theta)$=$v_{\theta}(r)\bm{t}(\theta)$ can be obtained as
\begin{equation}
\bm{v}(r,\theta) = \begin{cases}
    \frac{\omega}{2} \Bigl[1- \frac{(R-s)^2}{R^2} \Bigr]r \bm{t}(\theta)
 & (0 \leq r \leq R-s) \\
    \frac{\omega}{2} \Bigl(1- \frac{s}{R}\Bigr)^2\frac{R^2-r^2}{r} \bm{t}(\theta) & (R-s \leq r \leq R) \\
    0 & (r>R)
  \end{cases}
\end{equation}
where $\bm{t}(\theta)=(-\sin \theta, \cos \theta)$ is the unit orthoradial vector at the angular position $\theta$. The quantity $s$ is \SI{4.6}{\micro\meter} estimated from experimental data. In the following section, this analytic formulation is used to define the order parameter of AFMV pattern.

$ $

$\textbf{Order parameter of AFMV pattern}$

Here we show the derivation of order parameter of anti-ferromagnetic vortices (AFMV) pattern, given by Eq. (1) in main text. This order parameter compares the matching between the observed pattern of vortex pair in experiments and numerically calculated AFMV. For the numerical calculation of AFMV, the phenomenological description of vortex confined in boundary is considered as follows: for each circle composing the doublet microwell, we set an index $j$, 1 stands for the left side and 2 for the right side. We define two sets of polar coordinates $(r_j, \theta_j)$; one for left circle is $(r_1, \theta_1)$ and the other for right circle is $(r_2, \theta_2)$. The origin of $j$ polar coordinates is set at the center of $j$ circle. We consider $\bm{t}_j(\theta_j)$ the base polar orthoradial vector at the angular position $\theta_j$ centered on the center of the circle $j$ for $0\leq r_j \leq R$. In particular, we have $\bm{v}_j$$(r_j,\theta_j)=v_{\theta}(r_j)\bm{t}_j(\theta_j)$ where $v_{\theta}(r_j)$ is given by Eq. (15) and $\omega$ is the vorticity discussed at the previous section. 

We then consider vortices showing AFMV pattern in the doublet microwell. In addition to the boundary condition of a doublet of circles that is characterized by $R$ and $\Delta$, the polar coordinates $(r,\theta)$ is given to define the internal space. The origin of polar coordinates is placed at the centroid of the doublet shape. The velocity field, $\bm{v}$$(r,\theta)$, is in turn considered as the superposition of two vortices in $j$=1 and 2 circles. Because two vortices in AFMV pattern show opposite angular velocities, we can write $\bm{t}_1(\theta)=-\bm{t}_2(\theta)$ and then describe the velocity field as
\begin{equation}
\bm{v}(r,\theta) = \sum_{j} \bm{v}_j(r_j,\theta_j) =  \sum_{j} v_{\theta}(r_j) \bm{t}_j(\theta_j)  .
\end{equation}
The expected streamline of an AFMV pattern with a velocity field $\bm{v}(r,\theta)$ lies on the unit vector $\bm{u}(r,\theta)$ such that 
\begin{equation}
\bm{u}(r,\theta) =\frac{\bm{v}(r,\theta)}{|\bm{v}(r,\theta)|}   .
\end{equation}

To describe the transition between FMV and AFMV patterns, we consider the deviation from the expected AFMV pattern given by the product of expected velocity orientation map $\bm{u}(r,\theta)$ and the one measured experimentally $\bm{p}(r,\theta)$. The order parameter $\Phi$ is then defined as
\begin{equation}
\Phi=\vert \langle \bm{p}(r,\theta)\cdot \bm{u}(r,\theta) \rangle \vert
\end{equation}
where $\langle \cdot \rangle$ denotes the ensemble average over all sites in a doublet microwell. One can find $\bm{p}(r,\theta) \cdot \bm{u}(r,\theta) = \cos(\psi(r,\theta) - \psi^0(r,\theta))$ where $\psi(r,\theta)$ and $\psi^0(r,\theta)$ are the orientational angles of $\bm{p}(r,\theta)$ and $\bm{u}(r,\theta)$, respectively. When an actual AFMV pattern is recorded in $\bm{p}(r,\theta)$, $\Phi$ is close to 1, while for an FMV pattern, it is close to 0.

\end{document}